\documentclass[%
 reprint,
 amsmath,amssymb,
 aps,
pra,
]{revtex4-2}

\usepackage{graphicx}
\usepackage{dcolumn}
\usepackage{bm}

\usepackage{xcolor}

\begin{document}

\preprint{APS/123-QED}

\title{Field theoretic approach to photodetection\\via decaying oscillators}

\author{Samin Tajik}
\affiliation{Department of Physics and Astronomy, University of British Columbia, Vancouver, BC, Canada}

\author{William G. Unruh}
\email{unruh@physics.ubc.ca}
\affiliation{Department of Physics and Astronomy, University of British Columbia, Vancouver, BC, Canada}
\affiliation{Department of Physics and Astronomy, Texas A\&M University, College Station, TX, USA}

\date{\today}

\begin{abstract}
 We develop a field-theoretic model of photodetection in which the detector is represented by $N$ harmonic oscillators that are linearly coupled to the radiation field and decay into a set of output channels, where the energy absorbed from the field is converted into excitations of the oscillators and is then transferred into the output channels, producing a measurable detection of the signal that the input field carries. Starting from the system's Lagrangian, we derive the detector’s dynamics and the corresponding measurement statistics from quantum field correlations. We illustrate how coherent and squeezed inputs modify the counting statistics, and we identify the parameter ranges in which the detector approximates an ideal broadband photodetector versus a narrowband mode analyzer. We show that with appropriate tuning, by matching the couplings of the fields to the oscillators, we get the maximum signal in the output channels. For a two-mode squeezed vacuum as an input, we show how the quantum correlation between the modes can be detected in this model. 
\end{abstract}


\maketitle


\section{\label{sec:level1}Introduction }

Since the early days of quantum optics, researchers have introduced various photodetector models to bridge quantum field theory and measurable laboratory signals. In particular, Roy Glauber, in his foundational work on quantum optics in the 1960s, formulated a theory of optical coherence that uses different orders of field correlation functions to determine photodetection rates and joint detection probabilities \cite{Glauber1963OpticalCoherence,Glauber1963CoherentIncoherent,Glauber1963PhotonCorrelations}. In quantum field theory in curved spacetime,  and for non-inertial motion in flat spacetime, following the idea of ‘’particle being what particle detector detects,” the analogous role is played by particle-detector models such as the Unruh–DeWitt (UDW) detector, introduced by Unruh in 1976, where a simple quantum system (often idealized as a two-level atom) couples locally to a scalar field \cite{Unruh1976} \cite{BirrellDavies1982}. Pioneering work \cite{Unruh1976}, together with a series of subsequent works, has involved localized particle detectors that interact with a scalar quantum field on a fixed spacetime along a fixed trajectory, where the detector’s transition probabilities, which depend on the trajectory and the interaction between the field and the detector, characterize its response \cite{PhysRevD.29.1047}.
Moreover, Caves and collaborators in the late 1980s developed continuous descriptions of photodetection and propagating quantum optical fields, particularly in the context of traveling-wave and continuum quantum optics \cite{CavesCrouch1987} \cite{Blow1990}.

The central motivation for our work is the Unruh effect; the cornerstone of quantum field theory in curved spacetime, which states that the Minkowski vacuum is a thermal Kubo–Martin–Schwinger (KMS) state when analyzed with respect to the notion of time translation, defined by a one-parameter family of Lorentz boosts, where orbits of these Lorentz boost isometries correspond to a family of uniformly accelerating observers \cite{Unruh1976} \cite{Fulling1973} \cite{Davies1975} \cite{Wald1994}. 
Since the direct detection of the effect remains challenging in the current experimental setups, as the characteristic temperature scale is extremely small for achievable accelerations, in recent decades, quantum optical and optomechanical systems have been shown to provide promising settings as a novel paradigm for investigating Unruh-like effects and the correlations/entanglements that acceleration radiation carries
\cite{PhysRevLett.121.071301}. In this picture, a wide range of alternative settings, trajectories, and analogue spacetimes are considered, including circular acceleration, which, following the work of Bell and Leinaas \cite{BellLeinaas1983, Leinaas1998AcceleratedElectrons} and Unruh’s analysis of acceleration radiation for orbiting electrons \cite{Unruh1998OrbitingElectrons}, has been the subject of many recent studies \cite{Biermann2020UnruhCircular, Bunney2024ThirdSound, Gooding2020InterferometricUnruhBEC, Schutzhold2006UnruhElectrons, MartinMartinez2011BerryUnruh}. In particular, for the detection of this temperature, continuous quasiparticle detectors, lasers, storage rings, and orbiting-electron scenarios have been studied as promising candidates, where the detector’s internal degrees of freedom interact with quantum fluctuations in a manner that can be compared to Unruh-like excitation \cite{Crispino2008}  \cite{Unruh2022BlackHolesAccelerationTemp}. In parallel to alternative trajectories, analogue-gravity platforms have also opened a complementary route, where rather than accelerating charges in the vacuum, one engineers an effective relativistic field theory for quasiparticles (e.g., phonons) and couples the probes to those fluctuations \cite{Unruh1981Experimental} \cite{Barcelo2005AnalogueGravity} \cite{Retzker2008BECUnruhMethods} \cite{FedichevFischer2003PRL, Gooding2024NJP}. An example is the interferometric “continuous detector” program developed by Gooding et al., where, considering a laser coupled to a Bose–Einstein condensate, the probing field can function as an effective UDW detector, and the readout is performed interferometrically to detect the signal in the laser’s phase fluctuations sourced by condensate excitations \cite{Gooding2020InterferometricUnruhBEC}\cite{4yfh-tm4f}. 

Moreover, Scully, Svidzinsky and Unruh \cite{Scully2022PRResearch} have recently shown that, from the perspective of an accelerating atom in its ground state, an Unruh–Minkowski photon can be emitted dominantly into the left Rindler wedge, leaving the atom in an excited state. The excited atom then goes back to its ground state through the spontaneous emission of another photon dominantly into the right Rindler wedge but orthogonal (in the Klein Gordon norm) to the first photon. This process therefore generates entangled photon pairs in a two-mode squeezed state. These results have inspired discussions of detection schemes that go beyond direct photo-detection to study correlations in the radiation associated with acceleration.  Reznik et al. have previously demonstrated that two spatially separated localized quantum systems interacting locally with a quantum field can become entangled despite having no direct interaction with one another, and that harvested correlations are sufficiently strong to violate Bell inequalities \cite{Reznik2003, Reznik2005}. Hence, following the idea that, for detecting Unruh-like signals, entanglement and correlations carried by the fields, and other related horizon-induced phenomena, to be accessed in analogue systems, one often needs to go beyond direct photo-detection, we aim to propose another detection scheme by introducing a conceptual detector model designed to capture such field-theoretic signatures in a way that is both analytically solvable and aligned with interferometric and continuous-probe readout techniques.
In our detection model, unlike conventional localized particle-detector models, the detector includes $N$ harmonic oscillators at a single spatial point, interacting with an incoming scalar field, and subsequently decaying into output channel. These oscillators can be regarded as models of atoms which interact with the photons inside a photon detector. This simple model dates back to Planck's 1900-1901 work on blackbody radiation~\cite{Planck1900}. We partition these oscillators into $R$ groups, with $M$ oscillators in each and couple each group to one output channel, where the $M$ oscillators respond collectively to the field. The form of coupling between the oscillators and the fields/channels is derivative coupling, where the time derivatives of the fields are coupled to the oscillators. In our model, the collective response of the oscillators determines how the incoming field is transferred; since there is no direct measurement of the incoming photons, the output fields can provide access to different orders of correlations. In particular, we study how the number of oscillators affects the couplings and the transmission bandwidth. We calculate different orders of correlations between the channels to show how information carried by the incoming field can be accessed. We then aim to use different incoming scalar fields' state as the input to examine the signal and the correlations between output channels. We focus on establishing the conceptual and mathematical framework at the model level, and defer extensions to more detailed, platform-specific detector physics (including experimentally realistic couplings, geometry, and noise channels) to future work.

\section{Theoretical Model}\label{model}
We consider a massless scalar input field $\psi(t, x)$ in $1+1$ dimensions, which can carry the signal we want to measure, a set of $R$ output channel fields $\phi_r(t, y)$ (one field per channel, where the index $r=0, \ldots, R-1$ labels both an output field and its associated oscillator group), and $N$ harmonic oscillators of identical frequency $\Omega$. The oscillators are partitioned into $R$ disjoint groups of size $M$ (so $N=MR$, with $M$ oscillators per channel). Group $r$ contains harmonic oscillators with configuration variables $\left\{Q_{r,m}\right\}_{m=0}^{M-1}$, where $m=0, \ldots, M-1$ labels oscillators in each group. Throughout our calculations, we use natural units where $\hbar=c=K_B=1$ and the $(+,-)$ metric signature. The total action for our model can be written as:
\begin{equation}
S=S_\psi+\sum_{r=0}^{R-1} S_{\phi_r}+S_Q+S_{\mathrm{int}}
\end{equation}
with free actions for the scalar fields:
\begin{equation}
\begin{aligned}
& S_\psi=\frac{1}{2} \int d t d x\left[\left(\partial_t \psi\right)^2-\left(\partial_x \psi\right)^2\right] \\
& S_{\phi_r}=\frac{1}{2} \int d t d y\left[\left(\partial_t \phi_r\right)^2-\left(\partial_y \phi_r\right)^2\right]
\end{aligned}
\end{equation}
and the $N$ oscillators:

\begin{equation}
S_Q=\frac{1}{2} \int d t \sum_{r=0}^{R-1} \sum_{m=0}^{M-1}\left[\dot{Q}_{r, m}^2-\Omega^2 Q_{r, m}^2\right],
\end{equation}

where we have assumed: $\psi(t, \mathbf{x})=\psi(t, x) \quad($ no $y, z$ dependence $), \quad \phi_r(t, \mathbf{x})=\phi_r(t, y) \quad$ (no $x, z$ dependence), in which case $(\nabla \psi)^2=\left(\partial_x \psi\right)^2$ and $\left(\nabla \phi_r\right)^2=\left(\partial_y \phi_r\right)^2$. Upon interaction, the input field $\psi$ excites all oscillators $Q_{r,m}$ with uniform strength $\mu$ at the point $(x,y)=0$; i.e., we take all oscillators to be located at a single point in space. Each of the $M$ oscillators in the group $r$ couples to the output channel $\phi_r$ with strength $\lambda$. We therefore define the interaction action in the form of derivative coupling as

\begin{equation}
\begin{aligned}
S_{\mathrm{int}}=\int d t & {\left[\mu \dot{\psi}(t, 0) \sum_{r=0}^{R-1} \sum_{m=0}^{M-1} Q_{r, m}(t)\right.} \\
& \left.+\lambda \sum_{r=0}^{R-1} \dot{\phi}_r(t, 0) \sum_{m=0}^{M-1} Q_{r, m}(t)\right] .
\end{aligned}
\end{equation}

\begin{figure}
    \centering    \includegraphics[width=0.8\linewidth]{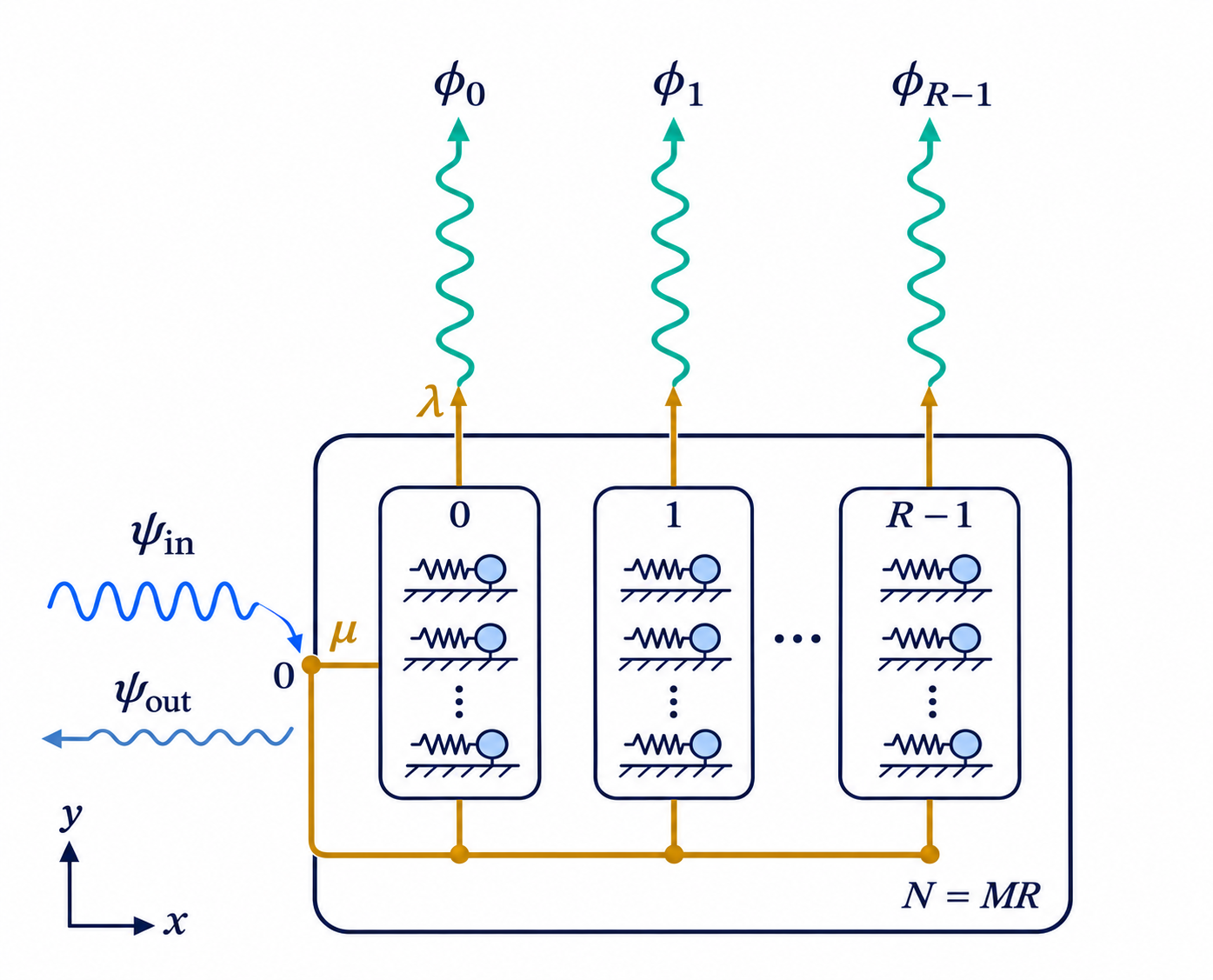}
    \caption{Schematic of the detector model. The incoming
field $\psi_{\mathrm{in}}$ propagates along $+x$ and couples
with strength $\mu$ to $N=MR$ harmonic oscillators.
The oscillators are grouped into $R$ groups of $M$
oscillators, each coupled with strength $\lambda$ to an
independent output field $\phi_r$ propagating along $+y$.
The reflected input field is denoted by
$\psi_{\mathrm{out}}$. All oscillators are localized at
$x=y=0$; their spatial separation in the diagram is
schematic.}
    \label{fig:placeholder}
\end{figure}

To simplify the system, and implement the reflecting boundaries, we impose a Neumann boundary condition on each of the field; on $\psi$ at $x=\epsilon{\rm~}$, and on each $\phi_r$ at $y=\epsilon$, noting that in the calculation we take $\epsilon \rightarrow0^+$ at the end, and that incoming $\psi$ propagates along $+x$, and the outgoing $\phi_r$ along $+y$. 

We define within each group $r$ the discrete Fourier modes:

\begin{equation}
Q_{r, \alpha}:=\frac{1}{\sqrt{M}} \sum_{m=0}^{M-1} e^{i \frac{2 \pi \alpha}{M} m} Q_{r, m}, \quad \alpha=0, \ldots, M-1 .
\end{equation}
Following from the discrete Fourier identity the uniform coupling within each group and the sum over $m$:

\begin{equation}
    \sum_{m=0}^{M-1} e^{i \frac{2 \pi \alpha}{M} m}= \begin{cases}M, & \alpha = 0, \\ 0, & \alpha \neq 0 ,\end{cases}
\end{equation} selects only the $\alpha=0$ (collective) Fourier mode,

\begin{equation}
\sum_{m=0}^{M-1} Q_{r, m}=\sqrt{M} Q_{r, 0},
\end{equation}
and the interaction reduces to: 
\begin{equation}
\mathcal{L}_{\mathrm{int}}=\mu \sqrt{N} \dot{\psi}(t, 0) Q_0(t)+\lambda \sqrt{M} \sum_{r=0}^{R-1} \dot{\phi}_r(t, 0) Q_{r, 0}(t),
\end{equation}

where $Q_0:=\frac{1}{\sqrt{N}} \sum_{j=0}^{N-1} Q_j$ is the entire system collective mode and $Q_{r, 0}$ is the group-  $r$ collective mode showing that while the input drives only $Q_0$ each output field couples to its own group coordinate $Q_{r,0}$

For our quantization we follow the conserved norm definition \cite{Unruh2011QuantumNI} \cite{Unruh2014} where for scalar fields $\psi, \phi_r $ and oscillator coordinates $Q_{r,m}$ with coupling constants $\lambda$ and $\mu$ can be written as

\begin{equation}
\begin{aligned}
\left\langle\Xi^{\prime}, \Xi\right\rangle=i & \left\{\int d x\left(\psi^{\prime *} \dot{\psi}-\dot{\psi}^{\prime *} \psi\right)\right. \\
& \quad+\sum_{r=0}^{R-1} \int d y\left(\phi_r^{\prime *} \dot{\phi}_r-\dot{\phi}_r^{\prime *} \phi_r\right) \\
& \quad+\sum_{r=0}^{R-1} \sum_{m=0}^{M-1}\left[Q_{r, m}^{\prime *}\left(\dot{Q}_{r, m}-\mu \psi(t, 0)-\lambda \phi_r(t, 0)\right)\right. \\
& \left.\left.\quad-\left(\dot{Q}_{r, m}^{\prime *}-\mu \psi^{\prime *}(t, 0)-\lambda \phi_r^{\prime *}(t, 0)\right) Q_{r, m}\right]\right\} .
\end{aligned}
\end{equation}

where $\Xi=\left\{\psi,\left\{\phi_r\right\},\left\{Q_{r, m}\right\}\right\}$ designates a complete classical solution of the equations of motion at any time $t$.

This norm is conserved by the equations of motion and relates the ingoing modes at $t \rightarrow-\infty$ to the outgoing modes at $t \rightarrow+\infty$. For a wave packet incident in the \(\psi\) channel, the conserved inner product reduces in the remote past to the free-field inner product in that channel. We use this asymptotic product to define the incoming annihilation operators.

Solving the linear system yields frequency-domain output fields. After quantizing the system, using the annihilation operator for the field $\psi$ defined as:

\begin{equation}
\hat{A}_j=i \lim _{t \rightarrow-\infty} \int_{-\infty}^0 d x\left[\psi_{j, \mathrm{in}}^* \partial_t \hat{\Psi}_{\mathrm{in}}-\left(\partial_t \psi_{j, \mathrm{in}}^*\right) \hat{\Psi}_{\mathrm{in}}\right]
\end{equation}

where $\psi_{j, \text { in }}$ are normalized incoming modes of the field $\psi$ and $\hat{\Psi}_{\text {in }}$ is the incoming field operator, so

\begin{equation}
\hat{\Psi}_{\mathrm{in}}(t, x)=\sum_j\left[\hat{A}_j \psi_{j, \mathrm{in}}(t, x)+\hat{A}_j^{\dagger} \psi_{j, \mathrm{in}}^*(t, x)\right].
\label{psiop}
\end{equation}

Similarly for the $\phi$ fields we write

\begin{equation}
\begin{aligned}
\hat{B}_{r j} & =i \lim _{t \rightarrow-\infty} \int_0^{\infty} d y\left[\phi_{r j, \mathrm{in}}^* \partial_t \hat{\Phi}_{r, \mathrm{in}}-\left(\partial_t \phi_{r j, \mathrm{in}}^*\right) \hat{\Phi}_{r, \mathrm{in}}\right], \\
&\hat{\Phi}_{r, \mathrm{in}}(t, y) =\sum_j\left[\hat{B}_{r j} \phi_{r j, \mathrm{in}}(t, y)+\hat{B}_{r j}^{\dagger} \phi_{r j, \mathrm{in}}^*(t, y)\right] ,
\label{phiop}
\end{aligned}
\end{equation}

where $j$ labels an incoming mode in channel $r$. Hereafter, hats on operators are omitted for notational simplicity, noting that mode functions and complex amplitudes remain ordinary c-number quantities. We define 

\begin{equation}
B_\beta(\omega)=R^{-1 / 2} \sum_r e^{2 \pi i \beta r / R} B_r(\omega)
\end{equation}

As shown in the appendix, \ref{app:classical} solving the coupled field and oscillators' equations of motion (EOM) with

\begin{equation}
\ddot{Q}_{r, m}+\Omega^2 Q_{r, m}=\mu \partial_t \psi(t, 0)+\lambda \partial_t \phi_r(t, 0) .
\end{equation}

for oscillators with a common frequency $\Omega$, with the Neumann boundary condition just behind the oscillators, and promoting the $\psi$, and $\phi$ field to operators using  \ref{psiop} and \ref{phiop} gives the following relations

\begin{equation}
\begin{aligned}
\tilde{A}(\omega) = &\left(1 +
\frac{2 i \omega N \mu^2}{
\Omega^2 - \omega^2 - i \omega (N \mu^2 + M \lambda^2)}\right) A(\omega) \\
& +\, 
\frac{2 i \omega \sqrt{N M} \lambda \mu}{
\Omega^2 - \omega^2 - i \omega (N \mu^2 + M \lambda^2)} B_0(\omega),
\label{Aop}
\end{aligned}
\end{equation}
and
\begin{equation}
\begin{aligned}
\tilde{B}_0(\omega) = &\left(1 +
\frac{2 i \omega M \lambda^2}{
\Omega^2 - \omega^2 - i \omega (N \mu^2 + M \lambda^2)}\right) B_0(\omega) \\
& +\, 
\frac{2 i \omega \sqrt{N M} \mu \lambda}{
\Omega^2 - \omega^2 - i \omega (N \mu^2 + M \lambda^2)} A(\omega),
\label{Bop}
\end{aligned}
\end{equation}

and 

\begin{equation}
\tilde{B}_\beta(\omega)=\left(1+\frac{2 i \omega M \lambda^2}{\left(\Omega^2-\omega^2\right)-i M \lambda^2 \omega}\right) B_\beta(\omega), \quad(\beta \neq 0),
\end{equation}

where we have expanded the input field $\psi(t, x)$ in incoming (right-moving) modes $e^{-i \omega(t-x)}$ with annihilation operator $A(\omega)$, and outgoing (left-moving) modes $e^{-i \omega(t+x)}$ with operator $\tilde{A}(\omega)$, and output fields $\phi_r(t, y)$ are expanded in the incoming modes along $-y$, $e^{-i \omega(t+y)}$, with their corresponding operator $B_r(\omega)$, and the outgoing modes along $+y, e^{-i \omega(t-y)}$, with operator $\tilde{B}_r(\omega)$. The annihilation operators of the outgoing channels $\phi_r$ are therefore a linear combination of the annihilation operators of the ingoing $\phi_r$ fields and the ingoing $\psi$ fields. Note that there is no mixing of creation and annihilation operators, i.e., no squeezing or amplification in the output (as opposed to the Unruh amplifier model; see \cite{Unruh2011QuantumNI}).

At resonance Eq~\ref{Aop} gives the input reflection amplitude $\left(M \lambda^2-N \mu^2\right) /\left(N \mu^2+M \lambda^2\right)$, which vanishes when $N \mu^2=M \lambda^2$, and Eq~\ref{Bop} under the same condition gives unit transmission from the input field, showing that if the right coupling condition is satisfied, the reflected part of the incoming field can go to zero, and all input amplitude is directed to the output. Assuming the incoming fields are not correlated, $\langle A^\dagger(\omega) B_0(\omega)\rangle=0$,
the mean spectral photon number in the collective ($\beta=0$) output mode is:
\begin{equation}
\begin{aligned}
\left\langle \tilde{B}_0^{\dagger}(\omega)\tilde{B}_0(\omega) \right\rangle
= &\, S_{B_0 A}(\omega)\,
\left\langle A^{\dagger}(\omega)A(\omega) \right\rangle \\
& +\, S_{B_0 B_0}(\omega)\,
\left\langle B_0^{\dagger}(\omega)B_0(\omega) \right\rangle ,
\end{aligned}
\end{equation}
with frequency dependent power transmission coefficients (transfer function):

\begin{equation}
S_{B_0 A}(\omega)=\left|\frac{2 i \omega \sqrt{N M} \lambda \mu }{\left(\Omega^2-\omega^2\right)-i \omega\left(N \mu^2+M \lambda^2\right)}\right|^2,
\label{SBA}
\end{equation}
and:
\begin{equation}
S_{B_0 B_0}(\omega)=\left|1+\frac{2 i \omega M \lambda^2}{\Omega^2-\omega^2-i \omega\left(N \mu^2+M \lambda^2\right)}\right|^2 .
\label{SBB}
\end{equation}
\begin{figure}
    \centering    \includegraphics[width=0.9\linewidth]{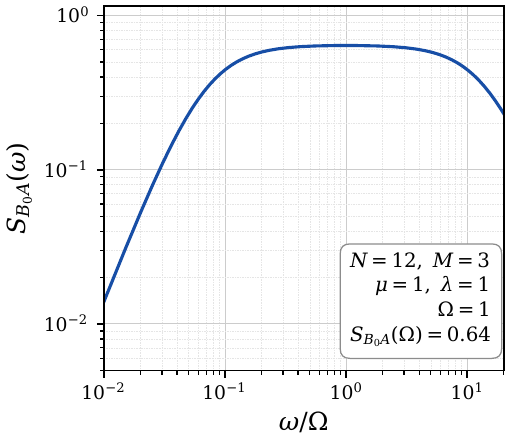}
    \caption{Power transmission coefficient $S_{B_0A}(\omega)$
from the input field to the collective output mode for
$N=12$, $M=3$, $\mu=\lambda=1$, and $\Omega=1$.
The collective couplings are mismatched,
$N\mu^2=12$ and $M\lambda^2=3$, giving a peak transmission
$S_{B_0A}(\Omega)=0.64$ at resonance.
Unity transmission is obtained at resonance when
$N\mu^2=M\lambda^2$.}

   \label{fig:S_BA}
\end{figure}

Suppose therefore that if prior to the interaction, the output channels are initially in vacuum, so that
$\langle B_0^\dagger(\omega)B_0(\omega)\rangle=0$, and the incident field in the input channel is prepared in a state $\rho_\psi$ with
$\langle A^\dagger(\omega)A(\omega)\rangle \equiv n_{\rm in}(\omega)$ then in spectrum becomes observable in the output channels via $S_{B_0 A} n_{\rm in}(\omega)$.

\section{Discussion}

\subsection{Power transmission bandwidth}

\begin{figure}
    \centering
    \includegraphics[width=0.9\linewidth]{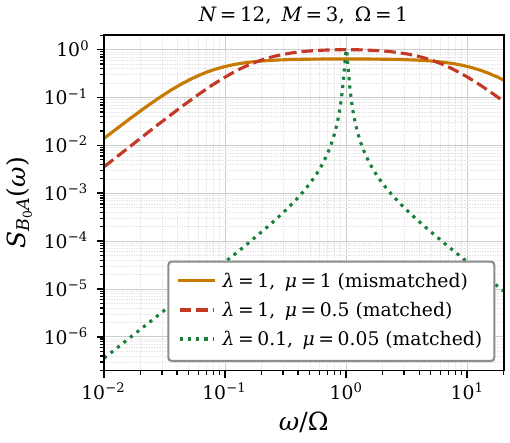}
    \caption{Comparison of power transmission for mismatched
and matched collective couplings, with $N=12$, $M=3$,
and $\Omega=1$. The solid curve corresponds to
$\lambda=\mu=1$ (mismatched), the dashed curve to
$\lambda=1$ and $\mu=0.5$ (matched), and the dotted curve
to $\lambda=0.1$ and $\mu=0.05$ (matched).
Matching, $N\mu^2=M\lambda^2$, eliminates input reflection
and gives unity transmission at resonance.
Reducing the collective couplings narrows the
transmission bandwidth.}
    \label{fig:compare}
\end{figure}

From \ref{SBA} we can see how the peak transmission efficiency depends on the relative strengths of the collective coupling $N\mu^2$ and $M\lambda^2$, and reaches unity when the right condition is satisfied; the transmission bandwidth is also set by their sum, i.e., $N\mu^2+M\lambda^2$, and in the weak coupling regime, $S_{B_0A}(\omega)$ exhibits a narrow resonance. For example, taking $\lambda=0.1$ (and $\mu$ fixed by the matching condition when applicable) with $N=12$ and $M=3$, the response reaches a sharp peak, as shown in
Fig.~\ref{fig:compare}.

\begin{figure}
    \centering
    \includegraphics[width=0.9\linewidth]{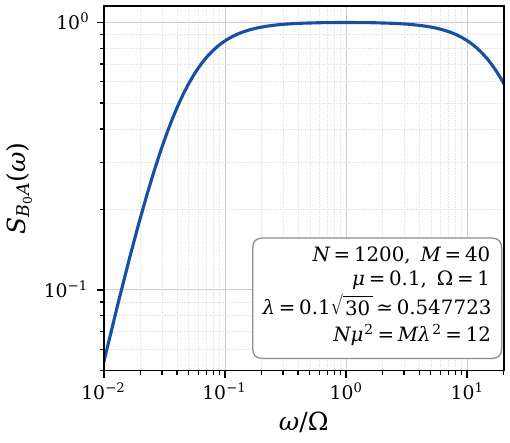}
    \caption{Broadband power transmission under collective
coupling matching. The parameters are $N=1200$, $M=40$,
$\mu=0.1$, $\lambda=0.1\sqrt{30}$, and $\Omega=1$,
so that $N\mu^2=M\lambda^2=12$.
The large collective damping
$N\mu^2+M\lambda^2$ produces a broad frequency interval
of near-unity transmission to the collective output mode.}
    \label{fig:largeN_bandwidth}
\end{figure}

However, one can see that increasing the array size broadens the effective linewidth, and for e.g. $N=1200$ and $M=40$,  (keeping $\mu=0.1$ while satisfying
$N\mu^2=M\lambda^2$,) the collective response recovers a broad and nearly flat passband,
as shown in Fig.~\ref{fig:largeN_bandwidth}.

What is particularly interesting about this model is that even in the weak-coupling regime, where the excitation probability of any single oscillator is small, the collective coupling produces a broad-bandwidth response. Hence, when the coupling matching condition $N \mu^2=M \lambda^2$ is satisfied, the reflection of the input field will vanish at resonance, and the power that is transferred to the output channels reaches unity; upon increasing the collective coupling, we can broaden the frequency range over which this remains close to unity. Note also that for large $N$, the probability that any one output channel contains even one photon goes to zero, and the fact that the model uses oscillators, rather than say two-level systems, become irrelevant, since each output channel only contains zero or one photon.

\subsection{Photon detection in output channels}

Following the Roy J. Glauber formulation of photodetection theory \cite{Glauber1963OpticalCoherence, Glauber1963PhotonCorrelations, Glauber1963CoherentIncoherent}, the first-order correlation function $G^{(1)}$ quantifies the field's coherence, and calculating it for our model reveals how well the incoming field amplitude in one output channel predicts the amplitude in another. Hence, it's the central quantity behind mean count rate, intensities, interference visibility, and (for stationary light) the power spectral density. Denoting by
$\Phi_r^{(\pm)}(t,y)$ the positive/negative-frequency parts of the output field
in channel $r$ (defined explicitly below), the first-order correlation function of the fields in two channels is
\begin{equation}
G^{(1)}(r_1,t_1,y_1;\,r_2,t_2,y_2)
=\left\langle \Phi_{r_1}^{(-)}(t_1,y_1)\,\Phi_{r_2}^{(+)}(t_2,y_2)\right\rangle .
\end{equation}
We also define the local intensity
\begin{equation}
\begin{aligned}
I(r, t, y) & :=G^{(1)}(r, t, y ; r, t, y), \\
I_i & :=I\left(r_i, t_i, y_i\right), \quad i=1,2, \ldots, R .
\end{aligned}
\end{equation}
so that the normalized degree of first-order coherence can be written as:
\begin{equation}
\begin{aligned}
g^{(1)}(r_1,t_1,y_1;\,r_2,t_2,y_2)
=\frac{G^{(1)}(r_1,t_1,y_1;\,r_2,t_2,y_2)}
{\sqrt{I_1 I_2}} .
\end{aligned}
\end{equation}

The frequency-domain quantized operators for the outgoing field are:
\begin{equation}
\tilde{B}_0(\omega)=\left(1+\kappa_1(\omega)\right) B_0(\omega)+\chi(\omega)\,A(\omega),
\end{equation}
and for $\beta\ge 1$,
\begin{equation}
\tilde{B}_\beta(\omega)=\left(1+\kappa_2(\omega)\right) B_\beta(\omega),
\end{equation}

where we used the shorthand notation:
\begin{equation}
    \kappa_1(\omega)= \frac{2 i M \lambda^2 \omega}{\left(\Omega^2-\omega^2\right)-i \omega\left(N \mu^2+M \lambda^2\right)}
\end{equation}
and

\begin{equation}
\chi(\omega)=\frac{2 i \omega \sqrt{N M} \mu \lambda}{\left(\Omega^2-\omega^2\right)-i \omega\left(N \mu^2+M \lambda^2\right)} ,
\end{equation}

and

\begin{equation}
\kappa_2(\omega)= \frac{2 i \omega M \lambda^2}{\left(\Omega^2-\omega^2\right)-i M \lambda^2 \omega}.
\end{equation}

The quantized positive-frequency part of the output field in continuous-mode representation for the channel $r$ is:

\begin{equation}
\Phi_r^{(+)}(t,y)=\int_0^{\infty}\frac{d\omega}{\sqrt{4\pi\omega}}\,
\widetilde{B}_r(\omega)\,e^{-i\omega (t-y)}.
\end{equation}

where we used transformation of the out field back to real-space channels $r$:

\begin{equation}
\widetilde{B}_r(\omega)=\frac{1}{\sqrt{R}}\sum_{\beta=0}^{R-1}
e^{-i 2\pi \beta r/R}\,\tilde{B}_\beta(\omega).
\end{equation}

It is convenient to split the field into contributions sourced by the input and by the
incoming channel operators, i.e
\begin{equation}
\Phi_r^{(+)}=\Phi_{r, A}^{(+)}+\Phi_{r, B}^{(+)} .
\end{equation}
The input-sourced contribution is
\begin{equation}
\Phi_{r, A}^{(+)}(t, y)=\frac{1}{\sqrt{R}} \int_0^{\infty} \frac{d \omega}{\sqrt{4 \pi \omega}} \chi(\omega) A(\omega) e^{-i \omega(t-y)}
\end{equation}
The channel contribution can be written using the following definition:

\begin{equation}
\begin{aligned}
\mathcal{B}_r(\omega):={}&(1+\kappa_1(\omega))\,B_0(\omega) \\
&+\sum_{\beta=1}^{R-1}e^{-i 2\pi \beta r/R}\,(1+\kappa_2(\omega))\,B_\beta(\omega)\ , 
\end{aligned}
\end{equation}

so that
\begin{equation}
\Phi_{r, B}^{(+)}(t, y)=\frac{1}{\sqrt{R}} \int_0^{\infty} \frac{d \omega}{\sqrt{4 \pi \omega}} \mathcal{B}_r(\omega) e^{-i \omega(t-y)}
\end{equation}

where the field operators have themselves been divided into their positive and negative frequency parts, i.e.

\begin{equation}
\Phi_r(t, y)=\Phi_r^{(+)}(t, y)+\Phi_r^{(-)}(t, y) .
\end{equation}

We take all incoming channels ${B}_\beta$ to be in the vacuum state. In this case,
inter-channel correlations in the outputs originate solely from the input operator ${A}(\omega)$
entering the collective mode ($\beta=0$).

For two channels $r$ and $s$ one gets using $\left\langle\widetilde{B}_r^{\dagger}(\omega) \widetilde{B}_s\left(\omega^{\prime}\right)\right\rangle=\mathcal{N}_{r s}(\omega) \delta\left(\omega-\omega^{\prime}\right)$

\begin{equation}
G_{r s}^{(1)}\left(u_1, u_2\right)=\int_0^{\infty} \frac{d \omega}{4 \pi \omega} \mathcal{N}_{r s}(\omega) e^{-i \omega\left(u_2-u_1\right)},
\end{equation}

where $u_i=t_i-y_i$. Note that this only applies to stationary fields with frequency-diagonal normal correlations. For stationary fields, the first-order correlation function is the Fourier transform of the weighted spectral coherence $\mathcal{N}_{r s}(\omega) /(4 \pi \omega)$. In particular, the local normally ordered field intensity is

\begin{equation}
I_r=\int_0^{\infty} \frac{d \omega}{4 \pi \omega} \mathcal{N}_{r r}(\omega).
\end{equation}

\subsection{Intensity correlations between channels and joint counting rate}

Another important question we address with our detector model is whether we can convert input-field correlations, such as entanglement in a two-mode squeezed state (e.g., in the case of an Unruh signal), into observable quantum correlations between localized physical systems. To measure the correlation carried by the incoming field, we use the correlation function defined for two separate output fields and show that the correlation between the channels appears as the incoming field acquires quantum correlations. We also want to emphasize that our model focuses on the field-theoretic transfer and readout of these correlations, without explicitly modeling the final irreversible registration of individual detection events.
To characterize correlations between output channels, we use Glauber's second-order correlation functions of the output fields. The second-order (intensity--intensity) correlation function
\begin{equation}
\begin{aligned}
G^{(2)}(r_1,t_1,y_1;\,r_2,t_2,y_2)
=\Big\langle\,
&\Phi_{r_1}^{(-)}(t_1,y_1)\,\Phi_{r_2}^{(-)}(t_2,y_2) \\
&\times \Phi_{r_2}^{(+)}(t_2,y_2)\,\Phi_{r_1}^{(+)}(t_1,y_1)
\,\Big\rangle ,
\end{aligned}
\end{equation}
with normalized form
\begin{equation}
\begin{aligned}
g^{(2)}(r_1,t_1,y_1;\,r_2,t_2,y_2)
=\frac{G^{(2)}(r_1,t_1,y_1;\,r_2,t_2,y_2)}{I_1 I_2} ,
\end{aligned}
\end{equation}

defined only where $I_1 I_2 \neq 0$, examines whether two of the initially independent channels become correlated through the vacuum fluctuation or the quantum state of the incoming field $\psi$ without any direct interaction between them. Usually we deal with statistically stationary fields in optics where the correlation functions are invariant under displacement of the time variable and can be written in terms of time differences $\tau=t_2-t_1$. For zero-mean Gaussian field states, Wick's theorem yields: 

\begin{equation}
    \begin{aligned} & G^{2}{\left(r_1, r_2, \tau\right)=}\left\langle\Phi_{r_1}^{(-)} \Phi_{r_1}^{(+)}\right\rangle\left\langle\Phi_{r_2}^{(-)} \Phi_{r_2}^{(+)}\right\rangle+\\& \left\langle\Phi_{r_1}^{(-)} \Phi_{r_2}^{(+)}\right\rangle\left\langle\Phi_{r_2}^{(-)} \Phi_{r_1}^{(+)}\right\rangle +\left\langle\Phi_{r_1}^{(-)} \Phi_{r_2}^{(-)}\right\rangle\left\langle\Phi_{r_2}^{(+)} \Phi_{r_1}^{(+)}\right\rangle \\ &=I_1 I_2+\left|G_{1,2}^{(1)}(\tau)\right|^2+|F(\tau)|^2,
    \end{aligned}
\end{equation}

where $G^{(1)}$ is the phase-insensitive coherence and
$F_{r s}(1,2):=\left\langle\Phi_r^{(+)}(1) \Phi_s^{(+)}(2)\right\rangle$ is the phase-sensitive coherence (pair). Below, we first present the results for a coherent input state with nonzero displacement, then introduce temporal modes and calculate the output correlations for coherent and two-mode-squeezed input states.

\subsubsection{Monochromatic coherent input}

In the framework of continuous-mode quantum theory, where one formulates field theories in $3$D infinite space and for a single input field (such as our input of a massless scalar field), continuum coherent states $|\{\alpha(\omega)\}\rangle$ are shown to be generated from the vacuum state $|0\rangle$ by a generalization of the standard displacement operator:

\begin{equation}
|\alpha (\omega)\rangle={D}(\alpha)|0\rangle,
\end{equation}

with 

\begin{equation}
{D}(\alpha)=\exp \left[\int d \omega\left[\alpha(\omega) {A}^{\dagger}(\omega)-\alpha^*(\omega) {A}(\omega)\right]\right],
\end{equation}

with the standard creation and annihilation operators $A(\omega)$. Many quantum optics experiments are performed using a light beam from a single-mode laser with a bandwidth much smaller than that of other components of the optical system. Such a ''single-mode" coherent light beam can be represented by a monochromatic coherent input with amplitude:

\begin{equation}
\alpha(\omega)=(2\pi\,\alpha_0)^{\frac{1}{2}}\,\delta(\omega-\omega_0),
\end{equation}
where $\alpha_0$ is the time-independent mean photon flux per unit time, and $\omega_0$ is its characteristic frequency, assuming the initial phase is $\theta=2\pi n$, and taking $\alpha(\omega)$ as the infinite-duration monochromatic limit, noting that we consider infinite-duration monochromatic limit.

We define the mean (classical) output field in channel $r$ (positive frequency) as: 
\begin{equation}
\mu_r(t,y):=\left\langle \Phi_r^{(+)}(t,y)\right\rangle
=\frac{\chi(\omega_0)\,\sqrt{\alpha_0}}{\sqrt{2 R \omega_0}}\,
e^{-i\omega_0 u}.
\end{equation}
Since a coherent state factorizes all normally ordered moments, the first-order coherence
factorizes as
\begin{equation}
\begin{aligned}
G^{(1)}(r,s;\tau)
&:=\left\langle \Phi_r^{(-)}(t,y)\,\Phi_s^{(+)}(t+\tau,y)\right\rangle\\
&=\frac{\alpha_0}{2R \omega_0}\,|\chi(\omega_0)|^2\,e^{-i\omega_0\tau}=\mu_r^*(t,y)\,\mu_s(t+\tau,y).
\end{aligned}
\end{equation}
The equal-time second-order correlation is
\begin{equation}
\begin{aligned}
G^{(2)}_{\rm coh}(r,s;t)
&:=\left\langle \Phi_{r}^{(-)}(t)\Phi_{s}^{(-)}(t)\Phi_{s}^{(+)}(t)\Phi_{r}^{(+)}(t)\right\rangle \\
&=|\mu_{r}(t)|^2\,|\mu_{s}(t)|^2.
\end{aligned}
\end{equation}

Therefore, the normalized second-order coherence satisfies
\begin{equation}
g_{r s}^{(2)}(t)=\frac{G_{\mathrm{coh}, r s}^{(2)}(t)}{I_r(t) I_s(t)}=1 .
\end{equation}
as expected for a monochromatic coherent input: the output intensities are Poissonian and the channels show no bunching or antibunching, i.e. no  excess normally ordered intensity correlation beyond the product of mean intensities $\left(g^{(2)}=1\right)$. 

\subsection{Temporal modes and correlation calculation}

To describe the temporal and spectral structure of the incoming field, here, we adopt the temporal-mode (TM) representation, in which the continuum of frequency modes is expanded in terms of a discrete orthonormal basis of wave packets \cite{RaymerWalmsley2020}. This representation provides a framework for studying how the detector’s frequency-dependent response is transferred into the output signals and their correlation functions, noting that for out model the detector transfer function $\chi(\omega)$, acts on each wave packet and affect the waveform on each output channel.

\subsubsection{Coherent state in temporal modes}
In temporal mode representation, a coherent state is defined as:

\begin{equation}
|\alpha\rangle_j=\exp \left(\alpha {A}_j^{\dagger}-\alpha^* {A}_j\right)|\mathrm{0}\rangle
\end{equation}

with the discrete annihilation and creation operators, defined as a linear superposition of monochromatic-mode annihilation operators:

\begin{equation}
{A}_j=\int d \omega u_j^*(\omega) A(\omega),
\end{equation}

noting that for a set of complete, orthogonal weight functions $u_j$, the new operators obey the standard commutation relations $\left[{A}_j, {A}_k^{\dagger}\right]=\delta_{j k}$. 

Here, orthogonality is defined by $\int \mathrm{d} \omega u_j^*(\omega) u_k(\omega)=\delta_{j k}$, and completeness by $\sum_j u_j^*(\omega) u_j\left(\omega^{\prime}\right)=\delta\left(\omega-\omega^{\prime}\right)$. Then each ${A}_j^{\dagger}$ creates one excitation in a particular TM, and the excitations in the field (modes) can be written as

\begin{equation}
\nu_j(t, x)=\int_0^{\infty} \frac{d \omega}{\sqrt{4 \pi \omega}} u_j(\omega) e^{-i \omega(t-x)}
\end{equation}

We will consider a coherent excitation of a single temporal mode $J$, whose spectral envelope is chosen to be Gaussian

\begin{equation}
u_J(\omega)=\mathcal{N} \exp \left[-\frac{T^2}{2}\left(\omega-\omega_c\right)^2\right],
\end{equation}

with $T>0$ being the parameter setting spectral width and $\omega_c$ as the central frequency, and  

\begin{equation}
\mathcal{N}=\left[\frac{\sqrt{\pi}}{2 T}\left(1+\operatorname{erf}\left(T \omega_c\right)\right)\right]^{-1 / 2}.
\end{equation}

We have ${A}_j|\alpha\rangle_j=\alpha |\alpha\rangle_j$, for the coherent state, so

\begin{equation}
|\alpha\rangle_J={D}_J(\alpha)|0\rangle, \quad\left\langle {A}_j\right\rangle=\alpha \delta_{j J} .
\end{equation}

The incoming field can be expanded in TM modes as

\begin{equation}
{\Psi}^{(+)}=\sum_j {A}_j \nu_j,
\end{equation}

and using the transformation in each channel we get

\begin{equation}
\Phi_r^{(+)}(t, y)=\sum_j {A}_j V_{r j}(t, y)+\Phi_{r, \text { vac }}^{(+)}(t, y) ,
\end{equation}

where

\begin{equation}
V_{r j}(t, y)=\frac{1}{\sqrt{R}} \int_0^{\infty} \frac{d \omega}{\sqrt{4 \pi \omega}} \chi(\omega) u_j(\omega) e^{-i \omega(t-y)}.
\end{equation}

We assume that all channels are in vacuum, and hence ${B}$ operators do not contribute to the correlation function. For the full basis state, we have

\begin{equation}
    |\psi\rangle=|\alpha\rangle_J \bigotimes_{j \neq J}|0\rangle_j,
\end{equation}

So, only mode $j=J$ is excited into coherent state, and all the other modes are in vacuum. Similarly

\begin{equation}
\Phi_s^{(-)}(t, y)=\sum_j {A}_j^{\dagger} V_{s j}^*(t, y)+\Phi_{s, \text { vac }}^{(-)}(t, y) .\end{equation}

Now we can calculate correlation functions. Glauber's first-order output correlation is

\begin{equation}
    G_{r s}^{(1)}(1,2)=\left\langle\Phi_r^{(-)}(t_1,y_1) \Phi_s^{(+)}(t_2,y_2)\right\rangle .
\end{equation}
Noting that $\left\langle\Phi_{\text {vac }}^{(-)} \Phi_{\text {vac }}^{(+)}\right\rangle=0$, we get

\begin{equation}
G_{r s}^{(1)}(1,2)=\sum_{j, k} V_{r j}^*(1) V_{s k}(2)\left\langle A_j^{\dagger} A_k\right\rangle .
\end{equation}

Since ${A}_k\left|\Psi_{\text {in }}\right\rangle=\alpha \delta_{k J}\left|\Psi_{\text {in }}\right\rangle,$ and $\left\langle\Psi_{\text {in }}\right| A_j^{\dagger}=\alpha^* \delta_{j J}\left\langle\Psi_{\text {in }}\right|,$ we obtain

\begin{equation}
  \begin{aligned}
\left\langle A_j^{\dagger} A_k\right\rangle & =\left\langle\Psi_{\text {in }}\right| A_j^{\dagger} A_k\left|\Psi_{\text {in }}\right\rangle \\
& =\alpha \delta_{k J}\left\langle\Psi_{\text {in }}\right| A_j^{\dagger}\left|\Psi_{\text {in }}\right\rangle \\
& =\alpha \delta_{k J} \alpha^* \delta_{j J} \\
& =|\alpha|^2 \delta_{j J} \delta_{k J} .
\end{aligned}  
\end{equation}
Substitution gives
\begin{equation}
G^{(1)}_{rs}(1,2)
=|\alpha|^2 V_{rJ}^*(1)V_{sJ}(2),
\label{eq:coherent-output-g1}
\end{equation}
or, explicitly,

\begin{equation}
\begin{aligned}
G^{(1)}_{rs}(1,2)
={}& \frac{|\alpha|^2\mathcal{N}^2}{R}
\int_0^\infty\!\!\int_0^\infty
\frac{d\omega_1\,d\omega_2}
     {4\pi\sqrt{\omega_1\omega_2}}
\\
&\times \chi^*(\omega_1)\chi(\omega_2)
\\
&\times
e^{-\frac{T^2}{2}
[(\omega_1-\omega_c)^2+(\omega_2-\omega_c)^2]}
\\
&\times
e^{i\omega_1(t_1-y_1)-i\omega_2(t_2-y_2)} .
\end{aligned}
\label{eq:Gaussian_G1}
\end{equation}

The output second-order correlation is

\begin{equation}
  G_{r s}^{(2)}(1,2)=\left\langle\Phi_r^{(-)}(1) \Phi_s^{(-)}(2) \Phi_s^{(+)}(2) \Phi_r^{(+)}(1)\right\rangle,  
\end{equation}

which gives

\begin{equation}
G_{r s}^{(2)}(1,2)=\sum_{j k l m} V_{r j}^*(1) V_{s k}^*(2) V_{s l}(2) V_{r m}(1)\left\langle A_j^{\dagger} A_k^{\dagger} A_l A_m\right\rangle .
\end{equation}

Therefore, the second-order correlation function reduces to

\begin{equation}
G_{r s}^{(2)}(1,2)=|\alpha|^4\left|V_{r J}(1)\right|^2\left|V_{s J}(2)\right|^2 .
\label{G2coh}
\end{equation}

Since $I_r(1)=|\alpha|^2\left|V_{r J}(1)\right|^2, \quad I_s(2)=|\alpha|^2\left|V_{s J}(2)\right|^2,$ we can also write \ref{G2coh} as

\begin{equation}
G_{r s}^{(2)}(1,2)=I_r(1) I_s(2) ,
\end{equation}

and the normalized second-order correlation is

\begin{equation}
g_{r s}^{(2)}(1,2)=\frac{G_{r s}^{(2)}(1,2)}{I_r(1) I_s(2)}=1,
\end{equation}

from which we have shown from which we have shown a coherent state of a single input TM produces $g_{rs}^{(2)}=1$ whenever the two intensities are nonzero, i.e. no excess intensity correlations between the channels.  

\subsubsection{Two-mode squeezed state in temporal modes}
From an accelerated observer's point of view, the Minkowski vacuum is described by a two-mode correlated state spanning both right and left Rindler wedges. It has been shown that the Minkowski vacuum entanglement between two Rindler wedges underlying the Unruh effect is analogous to two-mode squeezed light in quantum optics; for a uniformly accelerating atom in Minkowski spacetime, absorption of a Rindler particle corresponds to emission of a Minkowski particle. In this setting, the Unruh effect is understood as an accelerated atom, emitting a photon into a Minkowski vacuum via a transition from the ground state to an excited state\cite{PhysRevLett.91.243004, Svidzinsky2021PRResearch,Svidzinsky2021PRResearch,Svidzinsky2025PRD, Scully2018BlackHole}. This process also occurs for a two-level atom in a ground state held above the black hole event horizon, where it is first excited by emitting a negative-energy photon under the horizon and then spontaneously decays to the ground state by emitting a positive-frequency photon outside the horizon. Hence, a uniformly accelerated atom in Minkowski space-time emits entangled photon pairs in a squeezed state that mimics the entanglement of the Minkowski vacuum\cite{Scully2022PRResearch}. Here, we prepare the input by applying a two-mode squeezing operator to the vacuum, noting that the two modes are orthonormal wave packets of the same incoming field $\Psi$, and that both packets contribute to every output channel through the same frequency-dependent transfer function. We then compute the first- and second-order correlation functions for to examine our $N$-oscillators model as a detector of the correlation. In the quantum optics framework, this two-mode squeezed state is generated by the unitary squeezing operator \cite{scully1997quantum} acting on the two-mode vacuum state, and it corresponds to correlated photon pairs experimentally produced by techniques such as parametric down-conversion \cite{PhysRevLett.57.2520}.  In this state, the number of photons in the two modes is correlated, and we expect to detect this correlation in the output $\Phi_r$ channels.

Here, we construct the two-mode squeezed vacuum using the squeezing operator $S_{12}(\xi)$ based on our temporal mode formulation:

\begin{equation}
    |\xi\rangle=S_{12}(\xi)|0\rangle,
\end{equation}

where 

\begin{equation}
S_{12}(\xi)=\exp \left[\xi^* A_1 A_2-\xi A_1^{\dagger} A_2^{\dagger}\right], \quad \xi=r e^{i \theta},
\end{equation}

where the operators $A_1$ and $A_2$ annihilate two orthonormal temporal modes of the input field. Upon squeezing one gets

\begin{equation}
S_{12}^{\dagger} A_1 S_{12}=A_1 \cosh r-e^{i \theta} A_2^{\dagger} \sinh r ,
\end{equation}

and the same applies for $A_2$, giving

\begin{equation}
\begin{aligned}
& \left\langle A_1^{\dagger} A_1\right\rangle=\left\langle A_2^{\dagger} A_2\right\rangle=\sinh ^2 r, \\
& \left\langle A_1 A_2\right\rangle=\left\langle A_2 A_1\right\rangle=-e^{i \theta} \sinh r \cosh r , \\
& \left\langle A_1^{\dagger} A_2^{\dagger}\right\rangle=-e^{-i \theta} \sinh r \cosh r ,
\end{aligned}
\end{equation}

and

\begin{equation}
\left\langle A_1^{\dagger} A_2\right\rangle=0, \quad\left\langle A_1 A_1\right\rangle=\left\langle A_2 A_2\right\rangle=0.
\end{equation}

Note that the mean total input photon number is finite, and 

\begin{equation}
\left\langle N_{\text {in }}\right\rangle=2 \sinh ^2 r .
\end{equation}

Using the same definitions as for the coherent state calculation, the output field can be written as

\begin{equation}
\Phi_r^{(+)}(t, y)={A}_1 V_{r 1}(t, y)+A_2 V_{r 2}(t, y)+\Phi_{r, \text { vac }}^{(+)}(t, y) .
\end{equation}

Accordingly for the first order correlation function we get

\begin{equation}
G_{r s}^{(1)}(1,2)=\sum_{j, k=1}^2 V_{r j}^*(1) V_{s k}(2)\left\langle{A}_j^{\dagger} {A}_k\right\rangle .
\label{G1squeezed}
\end{equation}

Defining shorthand notation

\begin{equation}
n=\sinh ^2 r, \quad m=-e^{i \theta} \sinh r \cosh r,
\end{equation}

and using 

\begin{equation}
\left\langle A_j^{\dagger} A_k\right\rangle=n \delta_{j k}, \quad j, k \in\{1,2\} ,
\end{equation}

we can write Eq.\ref{G1squeezed} as

\begin{equation}
G_{r s}^{(1)}(1,2)=n\left[V_{r 1}^*(1) V_{s 1}(2)+V_{r 2}^*(1) V_{s 2}(2)\right] .
\end{equation}

Now, to look for the correlation between the channels, we calculate the second-order correlation function between the two output channels

\begin{equation}
G_{r s}^{(2)}(1,2)=\left\langle\Phi_r^{(-)}(1) \Phi_s^{(-)}(2) \Phi_s^{(+)}(2) \Phi_r^{(+)}(1)\right\rangle,
\end{equation}

which using Wick's theorem

\begin{equation}
\begin{aligned}
\left\langle A_j^{\dagger} A_k^{\dagger} A_l A_m\right\rangle= & \left\langle A_j^{\dagger} A_m\right\rangle\left\langle A_k^{\dagger} A_l\right\rangle \\
& +\left\langle A_j^{\dagger} A_l\right\rangle\left\langle A_k^{\dagger} A_m\right\rangle \\
& +\left\langle A_j^{\dagger} A_k^{\dagger}\right\rangle\left\langle A_l A_m\right\rangle,
\end{aligned}
\end{equation}

gives

\begin{equation}
\begin{aligned}
G_{r s}^{(2)}(1,2)= & n^2\left(\left|V_{r 1}(1)\right|^2+\left|V_{r 2}(1)\right|^2\right) \\
& \times\left(\left|V_{s 1}(2)\right|^2+\left|V_{s 2}(2)\right|^2\right) \\
& +n^2\left|V_{r 1}^*(1) V_{s 1}(2)+V_{r 2}^*(1) V_{s 2}(2)\right|^2 \\
& +n(n+1)\left|V_{r 1}(1) V_{s 2}(2)+V_{r 2}(1) V_{s 1}(2)\right|^2 .
\end{aligned}
\end{equation}

The normalized second-order correlations function, therefore gives

\begin{equation}
g_{r s}^{(2)}=1+\frac{\left|G_{r s}^{(1)}\right|^2}{I_r I_s}+\frac{\left|F_{r s}\right|^2}{I_r I_s} ,
\end{equation}

with

\begin{equation}
I_r(1)=n\left[\left|V_{r 1}(1)\right|^2+\left|V_{r 2}(1)\right|^2\right]
\end{equation}

which is related to the intensities in the channels, and

\begin{equation}
F_{r s}(1,2)=m\left[V_{r 1}(1) V_{s 2}(2)+V_{r 2}(1) V_{s 1}(2)\right] .
\end{equation}

which captures the contribution of anomalous pair correlations to the joint detection, and for two narrowband packets centered at $\omega_1$ and $\omega_2$ with $V_{r j}(u) \simeq v_{r j}(u) e^{-i \omega_j u}$, for $u_i=t_i-y_i$, where $v_{rj}$ includes the envelope and the transfer function $\chi$ we have

\begin{equation}
\begin{aligned}
F_{r s}(1,2) \simeq m[ & v_{r 1}\left(u_1\right) v_{s 2}\left(u_2\right) e^{-i\left(\omega_1 u_1+\omega_2 u_2\right)} \\
& \left.+v_{r 2}\left(u_1\right) v_{s 1}\left(u_2\right) e^{-i\left(\omega_2 u_1+\omega_1 u_2\right)}\right] ,
\end{aligned}
\end{equation}

showing that for two packets centered at $\omega_1$ and $\omega_2$ when two terms contribute their interference create a beat in the detection rate.

For the two-mode squeezed state as our input, therefore, the anomalous pair correlations contributing to the second order correlation function yields $g^{(2)}>1$ and becomes observable as correlations between the output channels.

\section{Conclusions}

In response to the question of going beyond direct detection schemes, in order to address the detection of " Unruh-like” signals, we developed a field-theoretic model of photodetection in which $N$ harmonic oscillators are coupled to an incoming scalar field signal and are then designed to decay into multiple groups of output channels. For this model, we derived the detector's dynamics and showed that collective oscillators’ modes, determining the response bandwidth, through matching the collective input and output couplings, $N\mu^{2}=M\lambda^{2}$, eliminate input reflection at resonance and yield complete signal transfer to the collective output mode, and that increasing the number of oscillators in the weak coupling regime broadens the response. We then adopted Glauber's formulation for the first- and second-order correlation functions of the field and calculated normally ordered field correlations to characterize the channel intensities and correlations that the incoming signal may carry.  For instance, we showed that for a coherent state as the input, the second-order correlation function factorizes into the product of mean intensities, giving the normalized function $g^{(2)}=1$ wherever the intensities are nonzero, and concluded that this accounts for no excessive correlation in the incoming field; whereas for a two-mode squeezed vacuum, anomalous pair correlations provide an additional contribution to the joint detection statistics, and hence account for the correlation in the incoming field’s state. Hence, our $N$-oscillator detector model can provide a series of readout channels with measurable intensity and coincidence statistics that determine the incoming field’s response, but it can also serve as a readout for the correlations carried by the incoming quantum field.

\begin{acknowledgments}

We thank Silke Weinfurtner, Jorma Louko, Cisco Gooding, and the
Gravity Lab group at the University of Nottingham for useful discussions. W.G.U. acknowledges partial support from the Hagler Institute for Advanced Study at Texas A\&M University. S.T. and W.G.U. acknowledge support from the Natural Sciences and Engineering Research Council of Canada (NSERC), which supported this research and provided doctoral stipend support for S.T. S.T. and W.G.U. gratefully acknowledge the support of the Science and Technology Facilities Council through Quantum Simulators for Fundamental Physics (ST/T006900/1), as part of the UKRI Quantum Technologies for Fundamental Physics programme, as well as support from the Leverhulme Research Leadership Award (RL-2019-020).

\end{acknowledgments}

\appendix*
\section{Solving the classical coupled boundary-value problem}
\label{app:classical}

In order to solve the classical equations of motion, we impose Neumann boundary condition at the origin, and use $x \leq 0, y \geq 0$, we have interaction is at $x=-\epsilon$, and output interactions are at $y=+\epsilon$, with $R$ output fields and $M$ oscillators per output field, so $N=M R$, where for the Lagrangian of our model we get

\begin{equation}
\begin{aligned}
& \square \psi+\mu \sum_{r, m} \dot{Q}_{r, m} \delta(x+\epsilon)=0, \\
& \square \phi_r+\lambda \sum_m \dot{Q}_{r, m} \delta(y-\epsilon)=0,
\end{aligned}
\end{equation}
where $\square=\partial_t^2-\partial_x^2$ for $\psi$ and $\square=\partial_t^2-\partial_y^2$ for $\phi_r$. The oscillator equation is

\begin{equation}
\ddot{Q}_{r, m}+\Omega^2 Q_{r, m}-\mu \dot{\psi}(t, -\epsilon)-\lambda \dot{\phi}_r(t, \epsilon)=0 .
\end{equation}

We then use the collective coordinate defined in the main text to write the equations in the mode space:

\begin{equation}
Q_{r, \alpha}=\frac{1}{\sqrt{M}} \sum_{m=0}^{M-1} e^{2 \pi i \alpha m / M} Q_{r, m}, \quad \alpha=0, \ldots, M-1 .
\end{equation}

which is collective mode within each group, noting that only $\alpha=0$ couples so:

\begin{equation}
Q_{r, 0}=\frac{1}{\sqrt{M}} \sum_{m=0}^{M-1} Q_{r, m},
\end{equation}

which gives:

\begin{equation}
\sum_m Q_{r, m}=\sqrt{M} Q_{r, 0} .
\end{equation}

Therefore, for oscillators we have:

\begin{equation}
\ddot{Q}_{r, 0}+\Omega^2 Q_{r, 0}=\sqrt{M} \mu \dot{\psi}(t, -\epsilon)+\sqrt{M} \lambda \dot{\phi}_r(t, \epsilon) .
\end{equation}

Now, transforming across the $R$ groups:

\begin{equation}
Q_\beta=\frac{1}{\sqrt{R}} \sum_{r=0}^{R-1} e^{2 \pi i \beta r / R} Q_{r, 0},
\end{equation}
 Hence, for the incoming field $\psi$ we get:

\begin{equation}
\square \psi+\sqrt{N} \mu \dot{Q}_0 \delta(x+\epsilon)=0,
\end{equation}

where

\begin{equation}
Q_0=\frac{1}{\sqrt{R}} \sum_r Q_{r, 0}=\frac{1}{\sqrt{N}} \sum_{r, m} Q_{r, m} .
\end{equation}

For the output channels, we also use the discrete Fourier transformation of the channels:

\begin{equation}
\phi_\beta=\frac{1}{\sqrt{R}} \sum_{r=0}^{R-1} e^{2 \pi i \beta r / R} \phi_r.
\end{equation}

Hence for the field in the output channels, we get:

\begin{equation}
\square \phi_\beta+\sqrt{M} \lambda \dot{Q}_\beta \delta(y-\epsilon)=0 .
\end{equation}

For the oscillators,

\begin{equation}
 \ddot{Q}_0+\Omega^2 Q_0=\sqrt{N} \mu \dot{\psi}(t, -\epsilon)+\sqrt{M} \lambda \dot{\phi}_0(t, \epsilon),   
\end{equation}

and

\begin{equation}
 \ddot{Q}_\beta+\Omega^2 Q_\beta=\sqrt{M} \lambda \dot{\phi}_\beta(t, \epsilon), \quad \beta \neq 0 .   
\end{equation}

Solving these equations for the fields $\phi_\beta$, assuming a Neumann boundary condition such that there is a mirror at $x = 0$ and $y = 0$, taking Fourier transform of the temporal part, and Laplace transform for the spatial part, gives

\begin{equation}
\begin{aligned}
\phi_\beta(t, y)= & \phi_\beta^H(t-y)+\phi_\beta^H(t+y) \\
& -\frac{\lambda \sqrt{M}}{2}\left[Q_\beta(t-|y-\epsilon|)+Q_\beta(t-y-\epsilon)\right] .
\end{aligned}
\end{equation}

Following the same steps for $\psi$ field, we get

\begin{equation}
\begin{aligned}
\psi(t, x)= & \psi^H(t-x)+\psi^H(t+x) \\
& -\frac{\mu \sqrt{N}}{2}\left[Q_0(t-|x+\epsilon|)+Q_0(t+x-\epsilon)\right] .
\end{aligned}
\end{equation}

We take the limit as $\epsilon\rightarrow0$, so that for
\begin{equation}
\begin{aligned}
\psi(t,-\epsilon)= & \psi^H(t+\epsilon)+\psi^H(t-\epsilon) \\
& -\frac{\mu \sqrt{N}}{2}\left[Q_0(t)+Q_0(t-2 \epsilon)\right] 
\end{aligned}
\end{equation}
and
\begin{equation}
\begin{aligned}
\phi_\beta(t, \epsilon)= & \phi_\beta^H(t-\epsilon)+\phi_\beta^H(t+\epsilon) \\
& -\frac{\lambda \sqrt{M}}{2}\left[Q_\beta(t)+Q_\beta(t-2 \epsilon)\right] ,
\end{aligned}
\end{equation}
we get

\begin{equation}
\begin{aligned}
& \psi(t, 0)=2 \psi^H(t)-\mu \sqrt{N} Q_0(t), \\
& \phi_\beta(t, 0)=2 \phi_\beta^H(t)-\lambda \sqrt{M} Q_\beta(t) .
\end{aligned}
\end{equation}

For the harmonic oscillators ($\beta \ne 0$) we have:

\begin{equation}
\ddot{Q}_\beta+M \lambda^2 \dot{Q}_\beta+\Omega^2 Q_\beta=2 \lambda \sqrt{M} \dot{\phi}_\beta^H ,
\end{equation}

and 

\begin{equation}
\begin{aligned}
\ddot{Q}_0+\left(N \mu^2+M \lambda^2\right) \dot{Q}_0+\Omega^2 Q_0= & 2 \mu \sqrt{N} \dot{\psi}^H \\
& +2 \lambda \sqrt{M} \dot{\phi}_0^H,
\end{aligned}
\end{equation}

where taking the Fourier transform and now taking the limit as $\epsilon\rightarrow 0$, and going to the Fourier transform space

\begin{equation}
\begin{gathered}
\psi^H(t)=\int d \omega a(\omega) e^{-i \omega t}, \quad \phi_\beta^H(t)=\int d \omega b_\beta(\omega) e^{-i \omega t} \\
Q_\beta(t)=\int d \omega \widetilde{Q}_\beta(\omega) e^{-i \omega t}
\end{gathered}
\end{equation}

gives

\begin{equation}
\widetilde{Q}_\beta(\omega)=-\frac{2 i \omega \lambda \sqrt{M} b_\beta(\omega)}{\Omega^2-\omega^2-i \omega M \lambda^2} .
\end{equation}
and:

\begin{equation}
\widetilde{Q}_0(\omega)=-\frac{2 i \omega\left[\mu \sqrt{N} a(\omega)+\lambda \sqrt{M} b_0(\omega)\right]}{\Omega^2-\omega^2-i \omega\left(N \mu^2+M \lambda^2\right) .}
\end{equation}

For the incoming wave we consider wavepacket in the $\psi$ channel, $\psi_H(t-x)=\int_{\omega>0} a_\omega e^{-i \omega(t-x)} d \omega$ with $a_\omega=0$ for $\omega<0$, in the limit as $t \rightarrow-\infty$, so that $\int \psi_H(t+x) \approx 0$ for all $x<0$

\begin{equation}
    \psi^H(x,t)=\int \tilde{a}(\omega) e^{-i \omega (t-x)} d \omega,
\end{equation}

and for each $\phi_\alpha$ incoming wave packet we consider

\begin{equation}
    \phi_\beta^H(t,y)=\int \tilde{b}_\beta(\omega) e^{-i \omega (t
    +y)} d \omega.
\end{equation}

For $t\rightarrow\infty$ the incoming wave packet converts itself to outgoing one. For the outgoing field we get

\begin{equation}
\begin{aligned}
\psi^{\text {out }}(t, x)=\int d \omega & {\left[\left(1+\frac{2 i \omega N \mu^2}{D(\omega)}\right) a(\omega)\right.} \\
& \left.+\frac{2 i \omega \sqrt{N M} \mu \lambda}{D(\omega)} b_0(\omega)\right] e^{-i \omega(t+x)} ,
\end{aligned}
\end{equation}

where we defined $D(\omega)=\Omega^2-\omega^2-i \omega\left(N \mu^2+M \lambda^2\right)$. Similarly for $\beta \ne 0$

\begin{equation}
\phi_\beta^{\text {out }}(t, y)=\int d \omega\left[1+\frac{2 i \omega M \lambda^2}{\Omega^2-\omega^2-i \omega M \lambda^2}\right] b_\beta(\omega) e^{-i \omega(t-y)},
\end{equation}

and for $\phi_0$ we have

\begin{equation}
\begin{aligned}
\phi_0^{\text {out }}(t, y)=\int d \omega & {\left[\left(1+\frac{2 i \omega M \lambda^2}{D(\omega)}\right) b_0(\omega)\right.} \\
& \left.+\frac{2 i \omega \sqrt{N M} \mu \lambda}{D(\omega)} a(\omega)\right] e^{-i \omega(t-y)}.
\end{aligned}
\end{equation}

Promoting the incoming and outgoing amplitudes to annihilation operators, we get the main-text relations for $\widetilde{A}, \widetilde{B}_0, \widetilde{B}_\beta$.

\nocite{*}

\bibliography{apssamp}

\end{document}